\documentstyle[multicol,aps,psfig]{revtex}
\begin{document}
\pagestyle{myheadings}
\title{Dynamics of Freely Cooling Granular Gases}
\author{Xiaobo Nie$^{1}$, Eli Ben-Naim$^{2}$, and Shiyi Chen$^{1,2,3}$} 
\address{
${}^{1}$Department of Mechanical Engineering, The Johns Hopkins University,
Baltimore, MD 21218\\
${}^{2}$Center for Nonlinear Studies and Theoretical Division,
Los Alamos National Laboratory, Los Alamos, NM 87545\\
${}^{3}$Peking University, Beijing, China}
\maketitle

\begin{abstract}

  We study dynamics of freely cooling granular gases in two-dimensions
  using large-scale molecular dynamics simulations. We find that for
  dilute systems the typical kinetic energy decays algebraically with
  time, $E(t)\sim t^{-1}$, in the long time limit. Asymptotically,
  velocity statistics are characterized by a universal Gaussian
  distribution, in contrast with the exponential high-energy tails
  characterizing the early homogeneous regime.  We show that in the late
  clustering regime particles move coherently as typical local
  velocity fluctuations, $\Delta v$, are small compared with the
  typical velocity, \hbox{$\Delta v/v\sim t^{-1/4}$}.  Furthermore,
  locally averaged shear modes dominate over acoustic modes.  The
  small thermal velocity fluctuations suggest that the system can be
  heuristically described by Burgers-like equations.
\vspace{2pt}
\\ {PACS:} 45.70.Mg, 47.70.Nd, 05.40.-a,  02.50.-r, 81.05.Rm
\end{abstract}
\begin{multicols}{2}
  Freely evolving granular media, i.e., ensembles of hard sphere
  particles undergoing dissipative inelastic collisions, exhibit
  interesting collective phenomena including clustering,
  vortices, shocks, and multiple dynamical regimes.  This dissipative
  nonequilibrium gas system can be described by kinetic theory in
  the early homogeneous phase where density fluctuations and velocity
  correlations are relatively small. However, quantitative
  characteristics of the late clustering regime such as the typical
  length and velocity scales, particle velocity distributions, 
  as well the corresponding continuum
  theory remain open questions
  \cite{Haff,Goldhirsch1,Goldhirsch2,McNamara1,McNamara2,Brey1,Deltour,Noije,Brito,Luding1,Luding2,Ben-Naim,Chen,Trizac,Orza,Esipov,Brey2,Brilliantov,Huthmann}.

  Recent molecular dynamics simulations in one-dimension have shown
  that the asymptotic energy decay is universal and that clustering
  corresponds to the formation of shocks in the inviscid Burgers
  equation \cite{Ben-Naim}. In this paper we study long time
  asymptotic dynamics of freely evolving granular gases in two
  dimensions using Molecular Dynamics (MD) simulations.  Treating the
  particles as identical, totally undeformable hard disks, we
  developed an efficient event-driven algorithm
  \cite{Alder,Rapaport,Marin} that allows us to probe the system well
  into the clustering regime. Our main result is that the asymptotic
  dynamics of dilute granular gases is universal as it is independent
  of the degree of dissipation.
  
In the simulations, $N=10^6$ identical disks of radius $R=0.15$ and
mass $m=1$ are placed in a two-dimensional system of linear dimension
$L=10^3$ with periodic boundary conditions.  The particle
concentration, $c=1$, so the mass density (or volume fraction) is
$\alpha=c\pi R^2=0.0707$. Initially, particles are distributed
randomly in space and their velocities are drawn from a Gaussian
distribution with zero mean and unit variance.  To ensure random
initial conditions, the system is first evolved under elastic
collisions only, with each particle undergoing $10^2$ collisions on
average. Then, time is reset to zero and each particle experiences up
to an average of $4 \times 10^4$ inelastic collisions.  In such a
collision the particle velocity, ${\bf v}$, changes according to
\begin{equation}
\label{rule}
  {\bf v}\to {\bf v}-\frac{1}{2}(1+r)({\bf g}\cdot {\bf n})\,{\bf n},
\end{equation}
with ${\bf g}={\bf v}-{\bf v'}$ the relative velocity of the colliding
particles and ${\bf n}$ a unit vector connecting the centers of the
two particles. In each collision, the normal component of the relative
velocity is reduced by the restitution coefficient, $0\leq r\leq 1$,
and the energy dissipation equals ${1\over 2}(1-r^2)({\bf g}\cdot{\bf
n})^2$. In the following, time is quoted in units of
$t_0=\frac{3}{20cR}\sqrt{m/E_0}$ which is proportional to the initial
mean free time, where $E_0$ is the average particle energy initially.

Inelastic collapse, the formation of a cluster via an infinite series
of collisions occurring in a finite time, poses a difficulty for
numerical simulations\cite{McNamara1}. To properly resolve such finite
time singularities, we take collisions to be perfectly elastic ($r=1$)
when the relative velocity falls below a pre-specified threshold ${\bf
g}\cdot {\bf n}<\delta$. This scheme, which mimics actual granular
particles where $r\to 1$ for sufficiently small relative velocities,
can be applied as long as the cutoff velocity $\delta$ is much smaller
than the root mean square (rms) velocity, $v_{\rm rms}\equiv\langle 
v^2\rangle^{1/2}$ \cite{Ben-Naim,Bizon1}. In our simulations, $v_{\rm
rms}>10^{-3}$, and $\delta=10^{-5}$.

Given the initial conditions, the collision sequence is deterministic,
and the system evolves freely (velocities are not controlled).  In the
absence of energy input, the system ``cools'' infinitely.  The average
particle energy, $E(t)={1\over 2}\langle v^2\rangle$, as a function of
time is shown in Fig.~1.  First, we confirmed that for sufficiently
small times, when the spatial distribution of particles remains
roughly uniform, the energy decay follows Haff's cooling law
\cite{Haff}
\begin{eqnarray}
E(t) = E_0\left(1+t/t_*\right)^{-2}.
\label{haff}
\end{eqnarray}
Here, \hbox{$(t_0/t_*)^{-1} = 3\sqrt{\pi}(1-r^2)g(2R)/20$}, where $g(r)$ is
the pair correlation function \cite{Luding1,Luding2}.

However, this cooling law eventually breaks down due to the formation
of dense clusters
\cite{Goldhirsch2,Deltour,Brito,Luding2,Ben-Naim,Chen,Orza}.  The
deviation from this law occurs at a time scale that ultimately
diverges when collisions become elastic.  Interestingly, our
simulations show that a universal decay law
\begin{eqnarray}
E(t) \simeq A(r) t^{-1},
\label{energy}
\end{eqnarray}
holds beyond this time scale. This implies that the typical velocity
decays with time according to \hbox{$v\sim t^{-1/2}$}. The typical
length scale explored by a particle, \hbox{${\cal L}(t)\sim \int_0^t
dt' v(t')\sim t^{1/2}$}, remains small compared with the system size,
${\cal L}\ll L$, and thus, finite size effects were negligible
throughout the simulations. The inset to Fig.~1 shows that the energy
decay law is independent of the cutoff velocity as long as $\delta
\leq 10^{-4}$ (we also verified that numerical errors were
irrelevant).

\bigskip
{\psfig{file=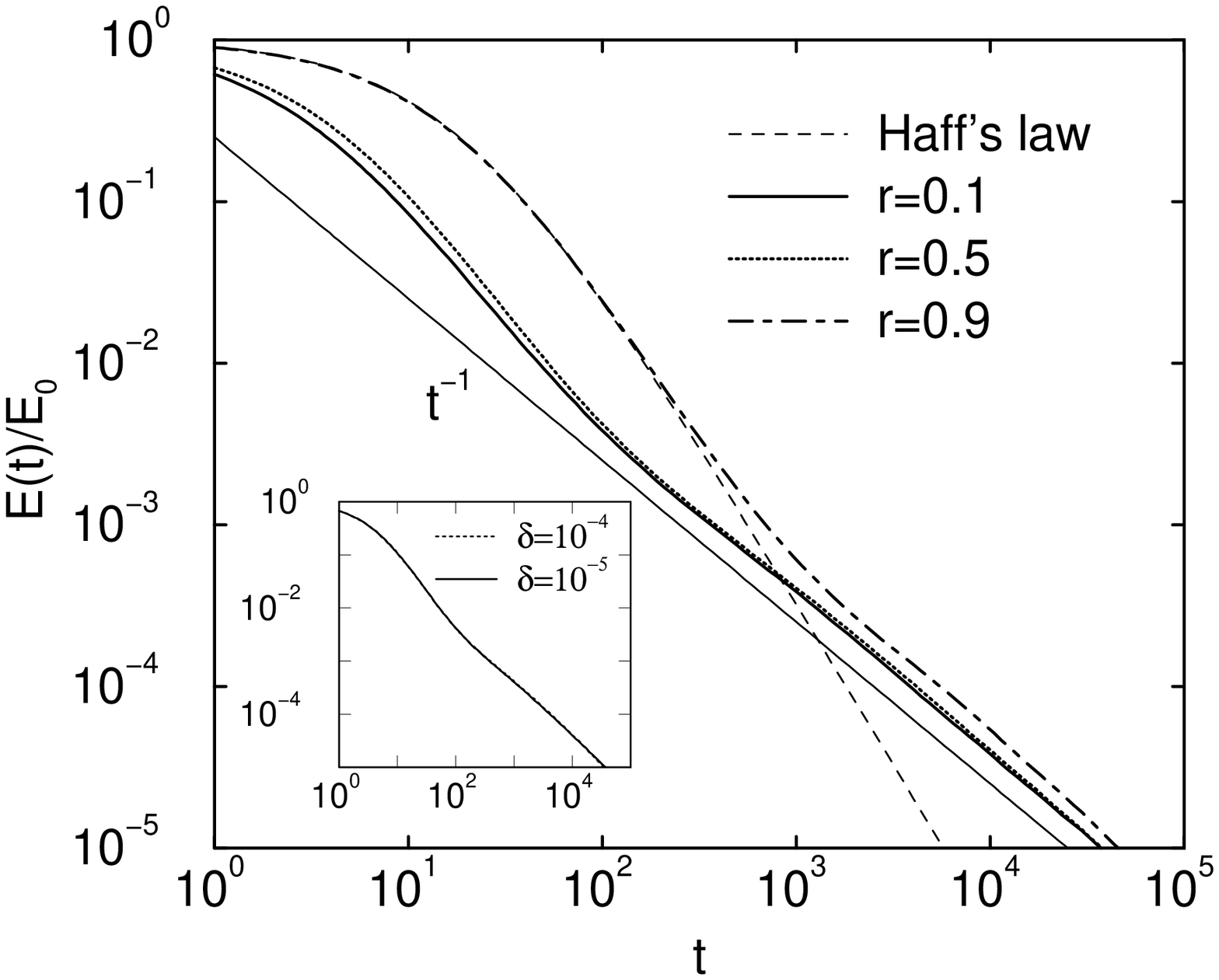,width=240pt}}
\noindent
{\small FIG.~1. The average particle energy $E(t)$ as a function of
time $t$ for three different restitution coefficients, $r=0.1$, $0.5$,
and $0.9$.  A line of slope $-1$ is shown for reference.  The inset
shows $E(t)$ as a function of $t$ for $r=0.1$ and different cutoffs
$\delta=10^{-4}$, $10^{-5}$.  Each line represents an average over
four independent realizations.}
\bigskip

Overall, this behavior is consistent with the $r$-independent energy
decay law, $E\sim t^{-2/3}$, found in one-dimension \cite{Ben-Naim},
although the exponent differs from the 1D value of 2/3.  This
asymptotic law (\ref{energy}) has not been observed in previous
numerical studies primarily due to insufficient temporal
range\cite{Goldhirsch2,Deltour,Brito,Luding2,Chen,Orza}.  For example,
power-law behavior was previously suggested by deformable sphere
molecular dynamics simulations in two- and
three-dimensions\cite{Chen}. However, the value of the scaling
exponent depended on parameters such as the system size and the
restitution coefficient.

The simulations show that as the volume fraction $\alpha$ was reduced,
the dependence of the prefactor $A(r)$ on the restitution coefficient
$r$ became weaker and weaker, suggesting completely universal behavior
in the dilute limit, $\alpha\to 0$. Further numerical studies with
smaller volume fractions are needed to fully resolve this issue.  If
indeed the prefactor $A$ is independent of the restitution coefficient
$r$, this implies a certain behavior of $t_c$, the time scale marking
the transition from the homogeneous to the clustering regime.
Matching the two asymptotic behaviors $E(t)\sim [(1-r)t]^{-2}$ for
$t\ll t_c$ with $E(t)\sim t^{-1}$ for $t\gg t_c$ shows that this
crossover time scale diverges according to $t_c\sim (1-r)^{-2}$ in the
quasi-elastic limit, $r\to 1$.

Next, we examined whether the entire velocity distribution, not merely
the typical velocity scale is independent of the restitution
coefficient. We studied $P(v,t)$, the probability distribution
function of the velocity magnitude $v\equiv |{\bf v}|$ at time 
$t$. Figure 2 clearly shows that a universal scaling function
underlies the velocity distribution when $t\gg t_c(r)$
\begin{equation}
P(v,t)\sim {1\over v^2_{\rm rms}}\Phi(z),\qquad z={v\over v_{\rm rms}}.
\end{equation}
This provides evidence that the asymptotic dynamics are universal.

The simulations suggest that the corresponding scaling function is
Gaussian, $\Phi(z)=\pi^{-1}\exp(-z^2)$, (see Fig.~2). We therefore
studied the kurtosis, particularly, the quantity $\kappa=\langle
v^4\rangle/\langle v^2\rangle^2-2$.  Initially, $\kappa = 0$, the
expected value for a Gaussian distribution in two-dimensions.  In the
intermediate regime, $\kappa $ is distinctly nonzero and strongly
depends on the value of the restitution coefficient. However, in the
asymptotic regime ($t\gg t_c$), $\kappa $ does not depend on
restitution coefficient and is very close to zero ($|\kappa|<0.1$), a
manifestation of the universal velocity statistics.  Given this
universality, we present data for a representative value of $r=0.1$ in
the rest of this paper.

\bigskip
{\psfig{file=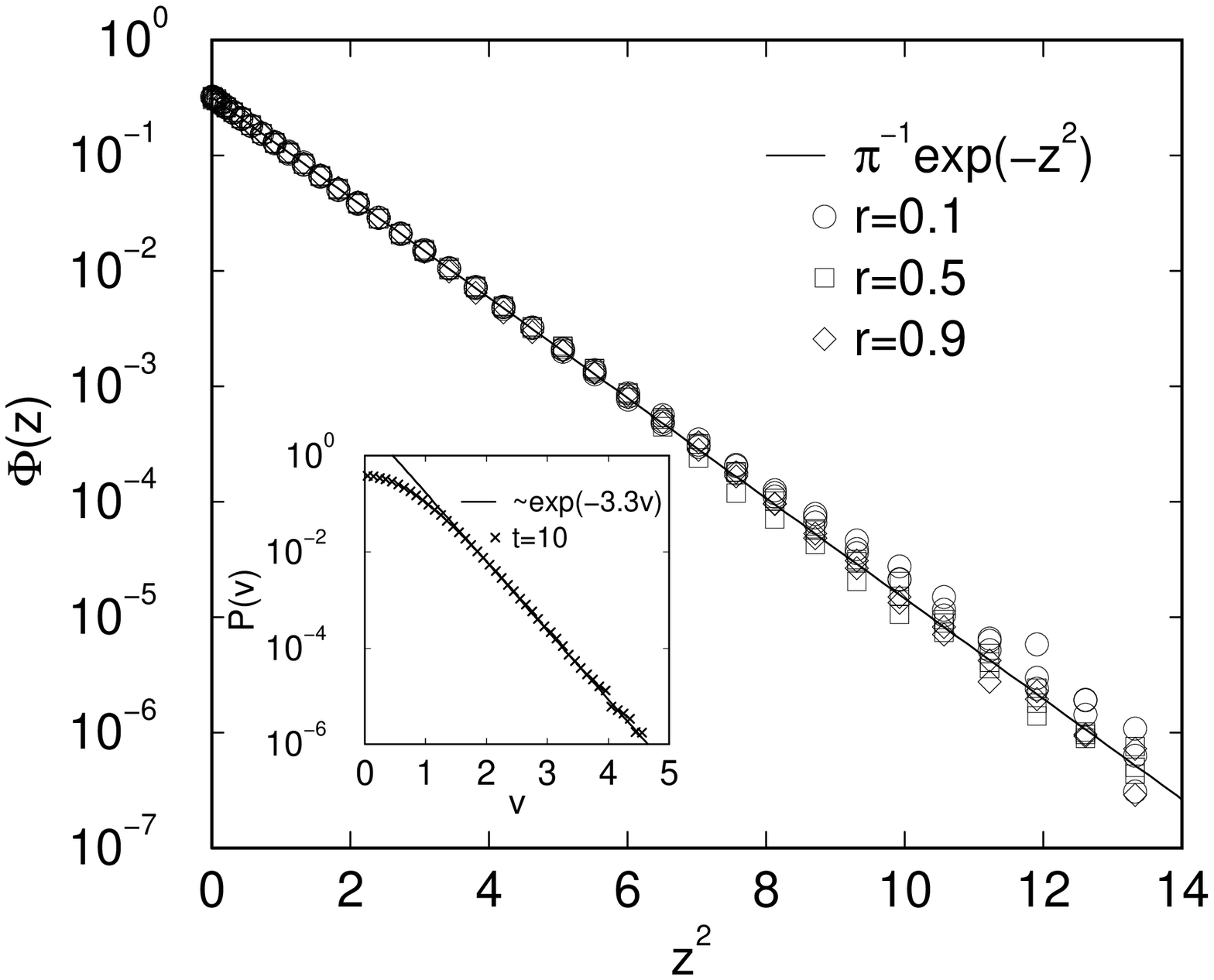,width=240pt}}
\noindent
{\small FIG.~2.  The scaling velocity distribution $\Phi(z)$ versus
the square of scaling variable $z=v/v_{\rm rms}$. The data corresponds
to three different restitution coefficients $r=0.1$, $0.5$, and $0.9$,
at three different times, $t=5 \times 10^2$ (excluding $r=0.9$), $ 2
\times 10^3$, $2 \times 10^4$, all in the clustering regime.  These
eight distributions follow a universal scaling function, a Gaussian
distribution.  The inset shows the exponential high-energy tail of
$P(v)$ at $t=10$ for $r=0.1$, where the velocity was rescaled by the
rms velocity.}
\bigskip

For completeness, we mention that in the intermediate homogeneous
regime, the distribution has an exponential high-energy tail, in
agreement with kinetic theory studies
\cite{Goldhirsch2,Noije,Esipov,Brey2,Brilliantov,Huthmann}. The
measured slope of $3.3$ is consistent with Direct Simulation Monte
Carlo (DSMC) \cite{Brey2} and MD simulations \cite{Huthmann} (see the
inset in Fig.~2).

\bigskip
{\psfig{file=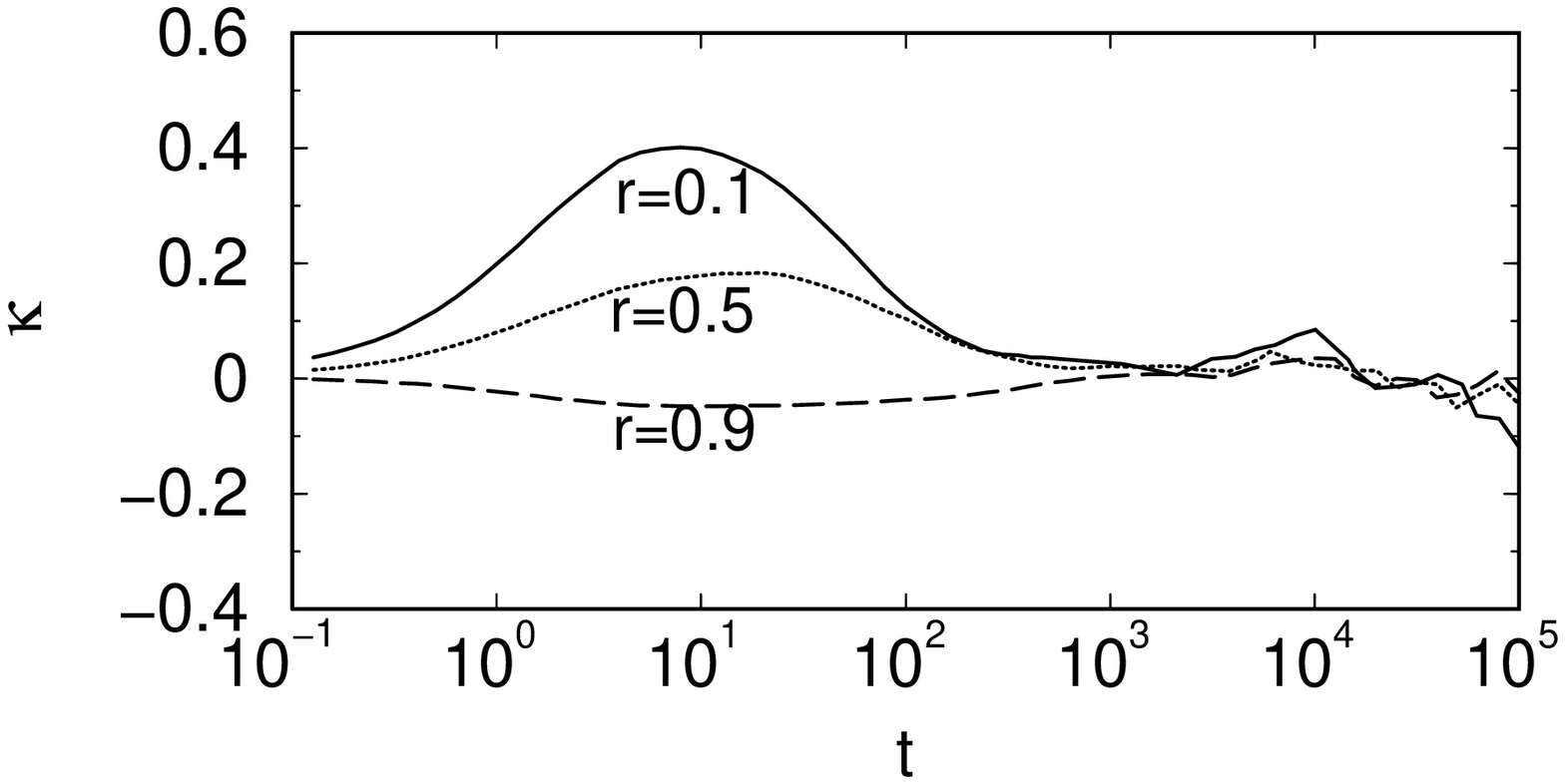,width=240pt}}
\noindent
{\small FIG.~3.  The adjusted kurtosis $\kappa$ as a function of time
for three restitution coefficients $r=0.1, 0.5$, and $0.9$.}
\bigskip

Both the Gaussian and the exponential extremal velocity statistics are
consistent with the following heuristic argument. The large-velocity
tails are dominated by the fastest particles ($v\approx 1$) initially
present in the system, still moving with their initial velocity at
time $t$. These particles managed to avoid any collisions with other
particles. The ``survival'' probability, $S(v,t)$, for such particles
decays exponentially with the volume they sweep till time $t$,
\hbox{$S(v=1,t) \propto \exp(-{\rm const.}\times t)$}.  Assuming a
stretched exponential large-velocity tail for the scaling function
\hbox{$\Phi(z)\sim \exp(-{\rm const.}\times |z|^{\gamma})$} as
$z\to\infty$ and the typical velocity decay $v \sim t^{-\beta}$ as
$t\to \infty$, and then matching the dominant behavior of the fastest
particles $P(1,t)=P(1,0)S(1,t)\propto \exp(-{\rm const.}\times t)$
with the expected scaling behavior $P(1,t)\propto
\Phi\left(t^{\beta}\right)\propto \exp\left(-{\rm const.}\times
t^{\beta\gamma}\right)$ yields the exponent relation $\beta\gamma=1$.
In the homogeneous regime, $\beta=1$ and thus $\gamma=1$, while in the
clustering regime, $\beta=1/2$ and thus, $\gamma=2$.

\bigskip
{\psfig{file=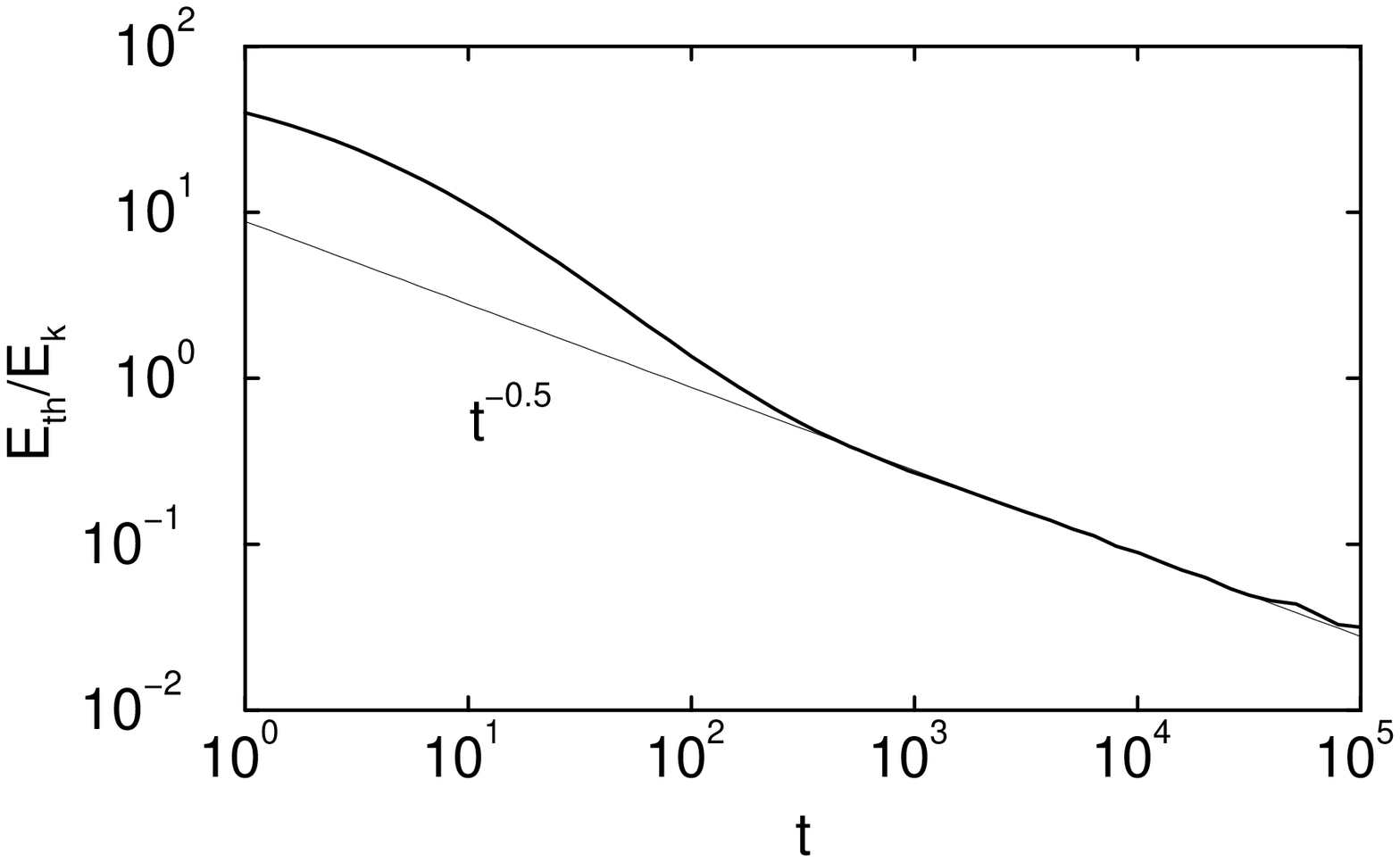,width=240pt}}
\noindent
{\small FIG.~4. The ratio of the thermal energy $E_{th}$ to the
kinetic energy $E_k$ as a function of time for $r=0.1$. A line of
slope $-1/2$ is also shown as a reference.}
\bigskip

To quantify collectives motions and the corresponding velocity
fluctuations in the clustering regime, we also measured the local
kinetic energy, $E_k$, and the local thermal energy, $E_{th}$.  The
local kinetic energy is defined by $E_k=\frac{1}{2} \langle\rho u^2
\rangle$ with the density $\rho$ and the velocity ${\bf u}\equiv (u_x,
u_y)$ obtained by averaging the corresponding quantities over a small
region of space.  The local thermal energy $E_{th}=\frac {1}{2}\langle
(v_x-u_x)^2+(v_y-u_y)^2\rangle$ is obtained by subtracting the average
local velocity ${\bf u}$ from the particle velocity ${\bf v}$ . Space
was divided into $128\times 128$ small boxes. As shown in Fig.~4,
initially, the thermal energy is large compared with the kinetic
energy, indicating that particles essentially move independently and
there are no collective motions.  In contrast, in the asymptotic
regime, $t\gg t_c(r=0.1)\approx 10^2$ the thermal energy becomes much
smaller than the kinetic energy, indicating that coherent motion of
particles becomes dominant.  The appearance of the collective motions
was suggested by linear stability analysis of the hydrodynamic
equations\cite{Goldhirsch1,Deltour} and observed in MD simulations
\cite{Goldhirsch1,Goldhirsch2,McNamara2,Deltour,Brito,Luding1,Luding2,Ben-Naim,Chen}.

Interestingly, we find that the ratio of thermal to kinetic energy
decays algebraically, $E_{th}/E_{k}\sim t^{-0.5}$, in the long time
limit.  The thermal energy, $E_{th}=\frac {1}{2}\langle (\Delta
v)^2\rangle$ quantifies local velocity fluctuations, $\Delta v$, and
using Eq.~(\ref{energy}),
\begin{equation}
 E_{th} \sim (\Delta v)^2\sim t^{-3/2}. 
\end{equation} 
In other words, local velocity fluctuations, $\Delta v\sim t^{-3/4}$
are small compared with the typical velocity, $v\sim t^{-1/2}$, as
$\Delta v/v\sim t^{-1/4}$. Thus, at least two distinct velocity scales
are needed to characterize particle velocities in the clustering
regime. Moreover, the relatively small velocity fluctuations imply
strong velocity correlations and a well-defined average local
velocity.  Thus, a hydrodynamic description is plausible in the
clustering regime.

\bigskip
{\psfig{file=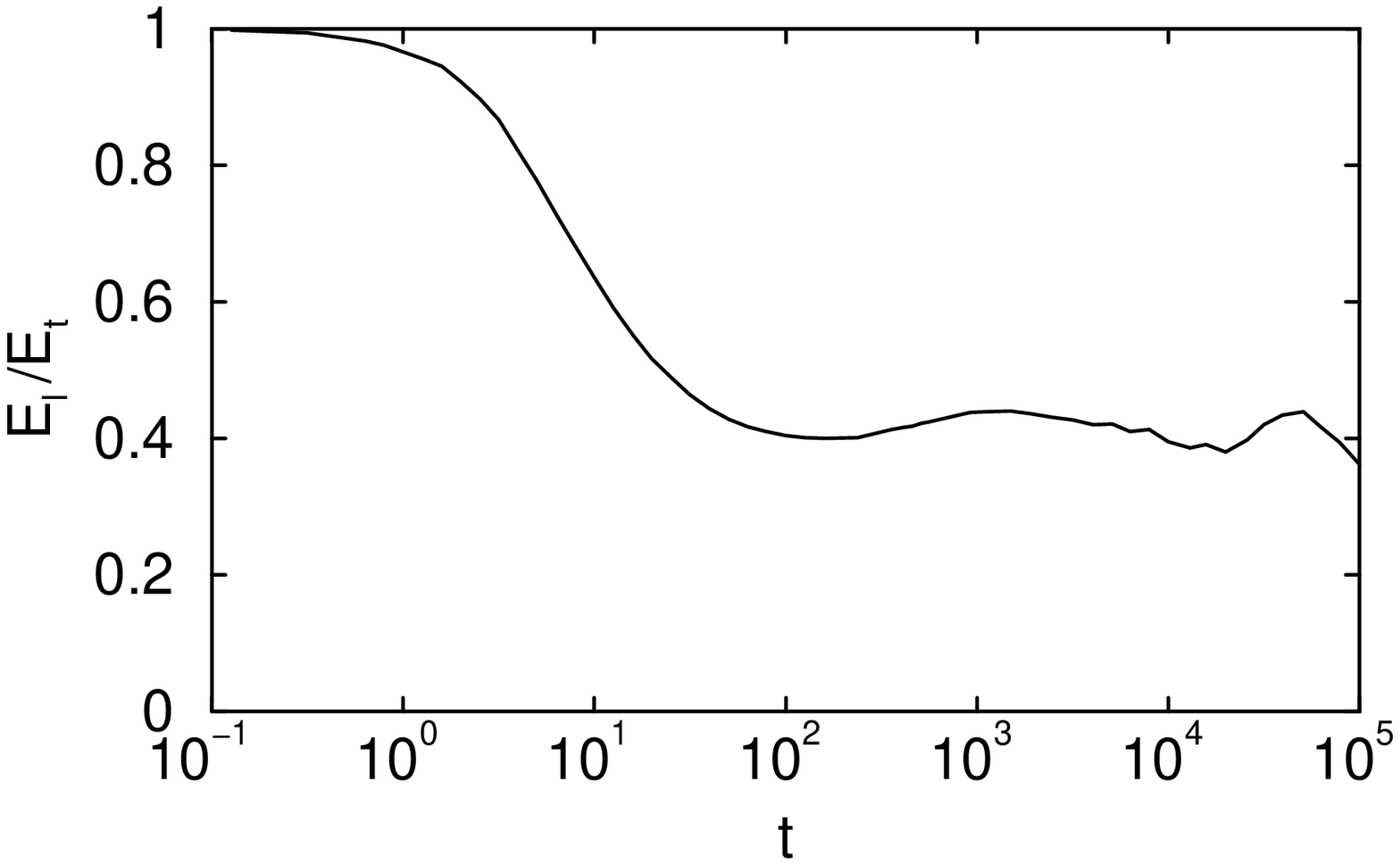,width=240pt}}
\noindent
{\small FIG.~5. The ratio of the longitudinal energy $E_l$ to the
transverse energy $E_t$ as a function of time.}
\bigskip

To distinguish between acoustic and shear modes, we transformed the
averaged velocity field into Fourier space and decomposed the velocity
as ${\bf u}({\bf k}) = {\bf u}_l({\bf k}) + {\bf u}_t({\bf k}) $.
Here ${\bf u}_t({\bf k})$ is the transverse velocity satisfying ${\bf
k} \cdot {\bf u}_t = 0$, and ${\bf u}_l({\bf k})$ is the longitudinal
velocity with $ {\bf k} \times {\bf u}_l =0 $. The corresponding
representation in physical space is $ {\bf u}(x,y)={\bf u}_l(x,y)+{\bf
u}_t(x,y)$, with $\nabla \cdot {\bf u}_t=0$, and $\nabla \times {\bf
u}_l=0 $.  The averaged longitudinal energy, i.e., the energy of
acoustic modes, is defined as the spatial average $E_l=\frac{1}{2}
\langle \rho u_l^2\rangle$, and the transverse energy, i.e., the
energy of shear modes, is defined as the space average of
$E_t=\frac{1}{2} \langle\rho u_t^2\rangle$. Initially, the system is
isotropic and consequently, the longitudinal and the transverse
energies are equal (see Figure 5). Since collective motions are
negligible initially, statistical fluctuations are isotropic as
well. The ratio of longitudinal to transverse energies decreases
steadily and eventually, it saturates at a value of roughly
$E_l/E_t\to 0.4$ for $t>t_c$. Therefore, shear modes dominate over
acoustic modes in the clustering regime. This behavior is consistent
with the formation of dense, thin, elongated clusters, where particles
move coherently, parallel to the cluster orientation
\cite{Goldhirsch1}. Interestingly, we did not observe obvious
correlations between the cluster size and its velocity.

Collective motion of granular media can be described by mass, momentum
and energy balance equations \cite{Goldhirsch1,Deltour,Jekins,Bizon2}.
We have seen that the thermal energy, i.e., the temperature $T$, is
negligible compared with the kinetic energy, and therefore, we may
expand the system around zero temperature. In particular, we can
ignore the pressure term in the momentum equation due to the fact that
$p \sim T$ while keeping the viscosity term since $\nu \sim
T^{1/2}$\cite{Deltour}.  The resulting governing equations are

\begin{eqnarray}
&\partial_t \rho + \partial_{\alpha} (\rho u_{\alpha})  =  0, \\
\label{density}
&\partial_t (\rho u_{\alpha}) + \partial_{\beta} (\rho u_{\alpha} 
u_{\beta})
 =  \partial_{\beta}\pi_{\alpha \beta}, \\
\label{mom}
&\pi_{\alpha \beta} = \partial_{\alpha} (\rho \nu u_{\beta}) +
 \partial_{\beta} (\rho \nu u_{\alpha}) -
 \partial_{\gamma} (\rho  \nu u_{\gamma}) \delta _{\alpha \beta},
\label{eos}
\end{eqnarray}
taken in the limit of vanishing viscosity, $\nu \rightarrow 0$.  The
above equations are similar to the two-dimensional Burgers equation
supplemented by the continuity equation, used to model large-scale
formation of matter in the universe\cite{Burgers,Shandarin}. Here, the
formation of shocks corresponds to dense, thin, string-like clusters
where particles move parallel to the cluster orientation. At least
qualitatively, this is consistent with the above findings that shear
modes dominate over acoustic modes.

In conclusion, our simulations show that in the low density and large
system limits, the kinetic energy of freely evolving granular media
decays with time as $t^{-1}$ in the long time limit. We have also
observed that the particle velocity distribution is Gaussian in this
asymptotic time regime. The above behavior is independent of the
degree of dissipation. We have studied the origin of this universality
and found that the asymptotic dynamics are dominated by collective
motions of particles or alternatively, shear modes.  In the clustering
regime, strong spatial velocity correlations develop, as local
velocity fluctuations are much smaller compared with the typical
velocity. We have also argued that the governing continuum equations
are similar to the two-dimensional Burgers equation with infinitely
small viscosity.

{\bf Acknowledgments.}  The simulations were performed on the Johns
Hopkins University cluster computer supported by the US NSF
(CTS-0079674).  This research was supported by China's NSF (10128204),
and the US DOE (W-7405-ENG-36)

\end{multicols}
\end{document}